\documentclass{llncs}
\usepackage{graphicx}
\usepackage{placeins}
\usepackage{float}
\usepackage{subfigure}
\usepackage{pdfpages}
\title{Reproducibility and Evolution of Diffusion MRI Measurements within the Cervical Spinal Cord in Multiple Sclerosis}

\author{Haykel Snoussi$^{1}$, Emmanuel Caruyer$^{2}$,  Beno\^\i t Comb\`es$^{1}$, Olivier Commowick$^{1}$, Elise Bannier$^{3}$, Anne Kerbrat$^{3}$, Julien Cohen-Adad$^{4}$, Christian Barillot$^{2}$}
\institute{Inria Rennes - Bretagne Atlantique, Rennes, 35042, France
            \and 
            IRISA-CNRS UMR 6074 VisAGeS Project-Team, Rennes, 35042, France
            \and
            Centre Hospitalier Universitaire de Rennes, Neurinfo platform, Rennes, France
            \and 
            Institute of Biomedical Engineering, Polytechnique Montreal, Montreal, QC, Canada}

\begin{document}
\maketitle

\begin{abstract}
In Multiple Sclerosis (MS), there is a large discrepancy between the clinical observations and how the pathology is exhibited on brain images, this is known as the clinical-radiological paradox (CRP). One of the hypotheses is that the clinical deficit may be more related to the spinal cord damage than the number or location of lesions in the brain. Therefore, investigating how the spinal cord is damaged becomes an acute challenge to better understand and overcome the CRP. Diffusion MRI is known to provide quantitative figures of neuronal degeneration and axonal loss, in the brain as well as in the spinal cord. In this paper, we propose to investigate how diffusion MRI metrics vary in the different cervical regions with the progression of the disease. We first study the reproducibility of diffusion MRI on healthy volunteers with a test-retest procedure using both standard diffusion tensor imaging (DTI) and multi-compartment Ball-and-Stick models. Then, based on the test re-test quantitative calibration,  we provide quantitative figures of pathology evolution between M0 and M12 in the cervical spine on a set of 31 MS patients, exhibiting how the pathology damage spans in the cervical spinal cord.

\end{abstract}

\section{Introduction}
Multiple Sclerosis (MS) is a neuro-inflammatory disease associated with a range of clinical symptoms and progressive physical disability. The use of non-invasive MRI techniques is key to a better understanding and follow-up of the pathology. However, there is usually a poor correlation between the radiological observation and the clinical outcome, something which is known as the clinical-radiological paradox (CRP). One of the potential improvements in our understanding of the pathology is using advanced quantitative MRI as well as investigate the extent of tissue damage in the cervical spinal cord \cite{barkhof2002clinico}.

Over the past decade, several groups started working on the improvement of MRI techniques for the spinal cord \cite{cohen2014quantitative}. Indeed, acquiring and processing MR images in spinal cord presents inherent challenges. Differences in magnetic susceptibility between soft tissues, air and bone make the magnetic field of spinal cord non-uniform and inhomogeneous. Also, given the small dimension of the cord cross-section (around 1 cm diameter at the cervical level), the specification and localization of lesions require a robust distinction between cerebrospinal fluid (CSF), grey matter (GM) and white matter (WM). In addition, besides the involuntary motion, acquiring MRI in the spine is hampered by the effect of cardiac and respiratory motion \cite{mohammadi2013impact,stroman2014current}.

Focal lesions are visible and detectable on conventional MRI (T1- and T2-weighted). However, more sophisticated MR imaging, namely diffusion MRI (dMRI), can provide quantitative information about tissue microstructure \emph{in vivo}, and therefore characterize axonal loss both diffuse and within the lesions \cite{clark2000diffusion}. Several metrics extracted from the diffusion MRI measurements are helpful as biomarkers of the pathology, such as the diffusion tensor imaging (DTI) characteristics: fractional anisotropy (FA); axial, radial and mean diffusivities (AD, RD and MD). Multi-compartment models also provide complementary measurements. In particular, using clinical data, it is possible to fit a Ball-and-Stick model \cite{behrens2007probabilistic}, from which one can extract the intrinsic diffusivity (ID), which is defined as the unique positive eigenvalue of the stick, as well as the free water weight (FWW). 

In this paper, we first investigate how reproducible these measures are for each vertebral level in the cervical spine, using a test-retest dataset on a group of 8 healthy subjects. We then compute these metrics on a group of 31 MS patients, and follow their longitudinal evolution between baseline and follow-up 12 months later. 

\section{Materials and Methods}
In this section we provide a description of the data acquisition, and of the image processing workflow for diffusion MRI analysis.

\subsection{Data acquisition}
\subsubsection{Patients and healthy volunteers}
Eight healthy volunteers (4 females, 4 males, median age 31 years, range 21-35) and 31 MS patients (21 females, 10 males, median age 30 years, range 20-49) were recruited in the study approved by the local research ethics committee. All participants provided informed written consent.

\subsubsection{MRI Acquisition}
MS patients and healthy volunteers were scanned on a 3T Siemens Verio scanner. Each subject was scanned twice with the same acquisition protocol. For MS patients, the second acquisition was performed within 12 months of the first one, however for healthy volunteers both acquisitions were performed few minutes apart. Thirty non-collinear diffusion-weighted images (DWI) were acquired at b = 900~s$\cdot$mm$^{-2}$, six non-DWI (b = 0) measurements and one non-DWI (b = 0) with an opposite phase encoding direction (PED) were also acquired. Scans were performed in sagittal orientation and head-feet (H-F) PED. The pulse sequence used for diffusion MRI is echo planar imaging (EPI). The reduced-FOV (field-of-view) technique was employed to reduce sensitivity of EPI to susceptibility artifacts. Sixteen slices were acquired with the following parameters without inter-slice gap: TR/TE = 3600/90~ms, with 2x2x2~mm$^3$ as the resolution, and image matrix 80x80. The total acquisition time for the dMRI sequence was approximately 7 minutes. The protocol also includes high-resolution T1-weighted image for anatomical reference with an isotropic 1x1x1~mm$^3$ resolution.

\subsection{Pre-processing and metrics extraction}
\subsubsection{Diffusion MRI pre-processing}
Motion between DWI were corrected using the method presented in \cite{xu2013improved} and implemented in the Spinal Cord Toolbox (SCT) \cite{de2017sct}. Then, dMRI data were corrected for susceptibility distortion using HySCO (Hyperelastic Susceptibility Artefact Correction) method as implemented in ACID-SPM toolbox presented in \cite{ruthotto2012diffeomorphic}. This method was recently shown to provide best results for distortion correction of spinal cord images \cite{snoussi2017comparison}. 

\subsubsection{Segmentation}
For each subject scan, whole spinal cord segmentation was carried out both on the mean DWI volume (b = 900~s$\cdot$mm$^{-2}$) and the T1-weighted using the Spinal Cord Toolbox (SCT) \cite{de2017sct}. A quality check was performed and parameters were modified, or manual adjustments were made when necessary.

\subsubsection{Computation of diffusion-based metrics}
We reconstructed Diffusion Tensor Images (DTI) and Ball-and-Stick models \cite{behrens2007probabilistic} in which the dMRI signal is split into a single isotropic component and a single anisotropic component. DTI was computed using SCT and Ball-and-Stick was reconstructed using in-house implementation. 

Metrics to be considered in the spinal cord quantification are: fractional anisotropy (FA),  mean diffusivity (MD), axial diffusivity (AD) and radial diffusivity (RD) for the DTI model, and intrinsic diffusivity (ID) -- defined as the diffusivity of the stick, and free water weight (FWW) for the Ball-and-Stick model. The objective to quantify these metrics is to test the presence of WM abnormalities in MS patients.

\subsubsection{Template-based analysis}
\begin{figure}
\begin{center}
  \includegraphics[width=.9\textwidth]{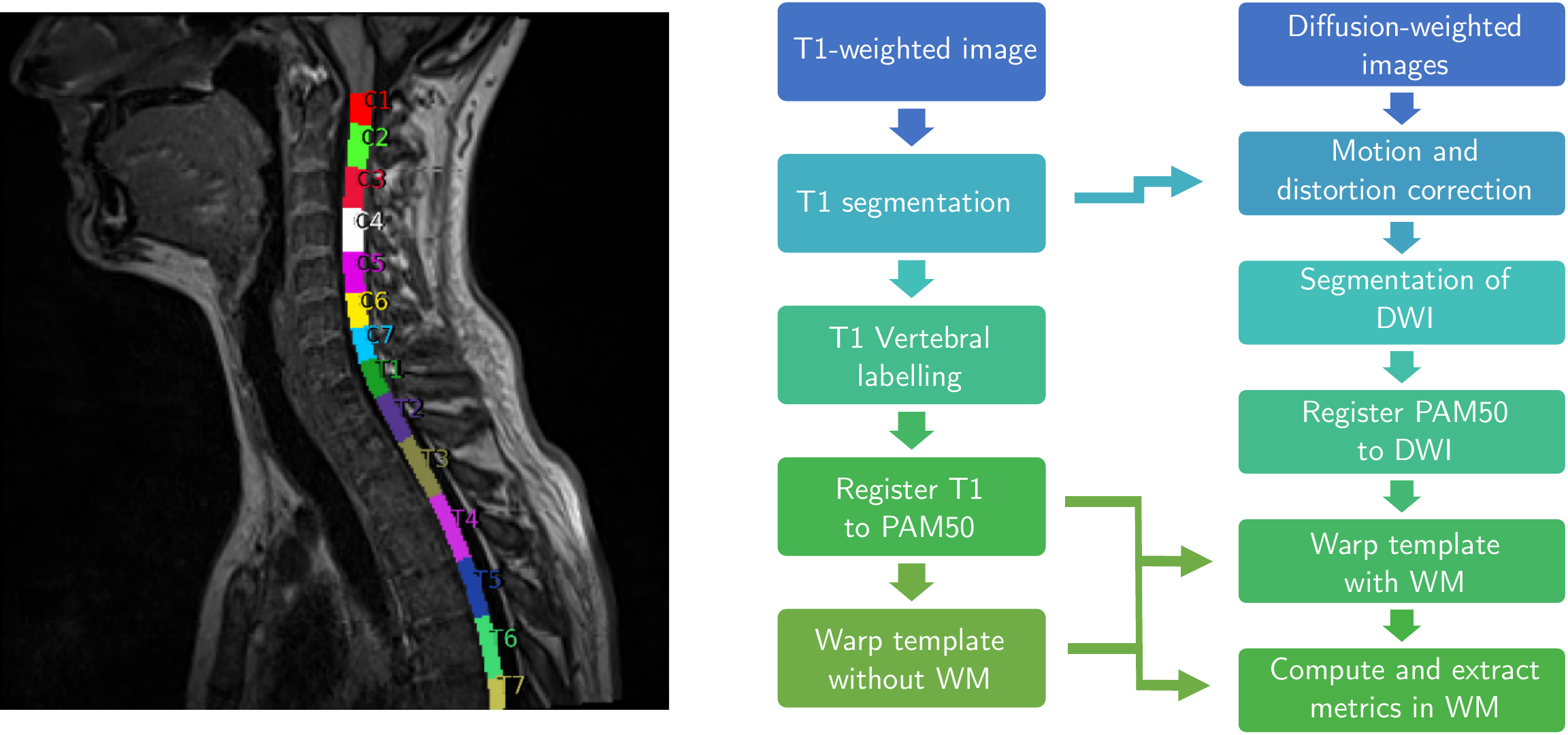}
  \caption{Left: T1 image with vertebral labels; right: imaging and processing pipeline for diffusion MRI and T1 images.\label{fig:pipeline}}
\end{center}
\end{figure}

Next, DWI data were registered to the PAM50 spinal cord template \cite{de2018pam50}, using a various affine and homeomorphic transformation between the mean of the DWI, the T1-weighted anatomical data and PAM50 template \cite{de2017sct}. Alignment with the template provides robust definition of the inter-vertebral levels for the spine. This enables computation of the average metrics in spinal cord using the atlas-based approach introduced in \cite{levy2015white}, which overcome biases related to partial volume effects. Compared to ROI and tractography approaches, this approach is less sensible to susceptibility distortions. DTI and  Ball-and-Stick metrics were estimated using a maximum a posteriori method. As a result, we can quantify diffusion-based metrics averaged for each inter-vertebral level between C1 and C7 within white matter. The processing pipeline as a whole is summarized in Fig.~\ref{fig:pipeline}.

\section{Results}

\subsection{Inter-subject and intra-subject variability on controls}
\begin{figure}[h]
\begin{center}
  \includegraphics[width=0.49\textwidth]{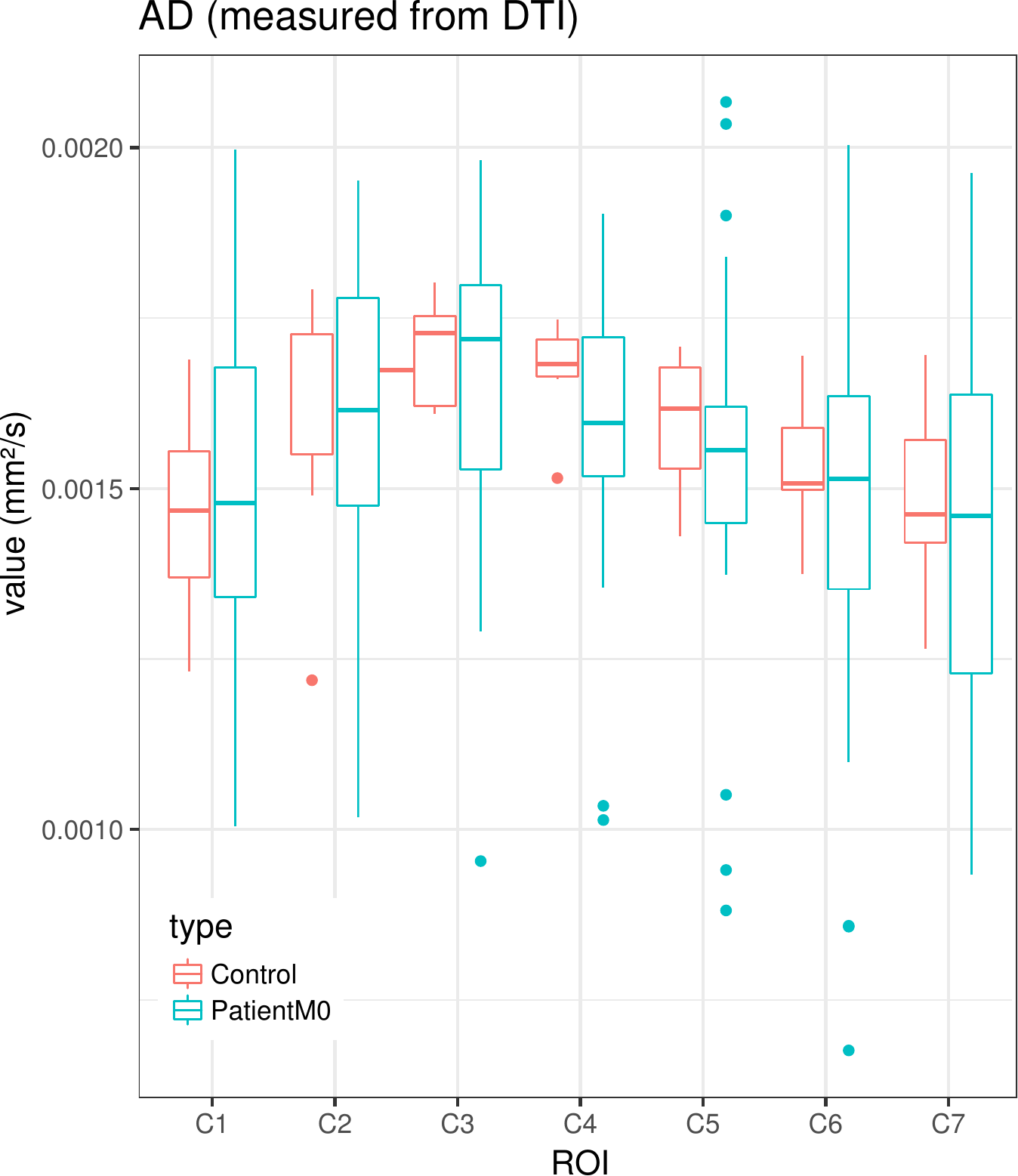}
  \includegraphics[width=0.49\textwidth]{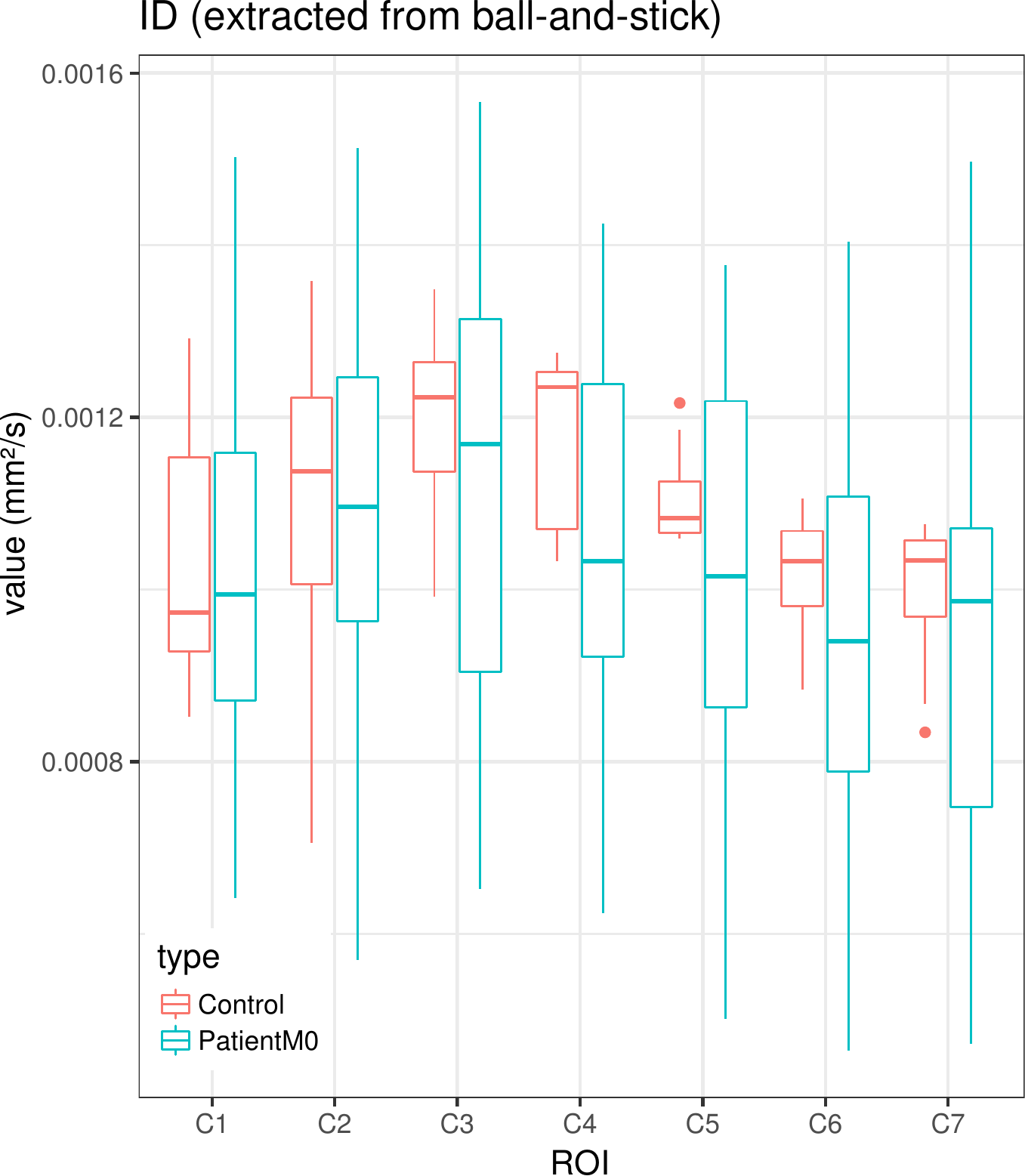}
  \caption{Distribution of AD (extracted from DTI) and ID (extracted from Ball-and-Stick) for controls and patients at M0.
  \label{fig:distribution-ad-id}}
\end{center}
\end{figure}

The variance across subjects of every metric was computed for each vertebral level in controls and in patients. As reported in Fig.~\ref{fig:distribution-ad-id} and Table~\ref{table:stdev}, the variance of almost every metric is higher in vertebral levels C1-C2 and C6-C7 in controls. This can be explained by the fact that larger distortions are observed in images at the top and the bottom of the field of view. In the following, we propose to use C3-C5 levels to extract averaged metrics with low cross-subject difference.

\begin{table}
\begin{center}
\begin{tabular}{c|c|c|c|c|c|c|c|c|c}
  \hline
  \multicolumn{9}{c}{\textbf{Levels}}\\
  \hline
  \textbf{Metrics} ~&~ C1 ~&~ C2 ~&~ C3 ~&~ C4 ~&~ C5 ~&~ C6 ~&~ C7 ~&~ C1C7 ~&~ C3C5 \\
  \hline
  AD ~&~ 0.20 ~&~ 0.21 ~&~ 0.09 ~&~ 0.10 ~&~ 0.12 ~&~ 0.12 ~&~ 0.23 ~&~ 0.07 ~&~ 0.07 \\
  FA ~&~ 82.38 ~&~ 54.59 ~&~  48.99 ~&~ 49.81 ~&~ 86.09 ~&~ 121.79 ~&~ 122.03 ~&~ 43.38 ~&~ 44.46 \\
  RD ~&~ 0.16 ~&~ 0.11 ~&~ 0.10 ~&~ 0.09 ~&~ 0.13 ~&~ 0.19 ~&~ 0.19 ~&~ 0.07 ~&~ 0.08 \\
  MD ~&~ 0.16 ~&~ 0.12 ~&~ 0.08 ~&~ 0.08 ~&~ 0.12 ~&~ 0.15 ~&~ 0.19 ~&~ 0.06 ~&~ 0.07 \\
  \hline
  ID ~&~ 0.17 ~&~ 0.22 ~&~ 0.14 ~&~ 0.15 ~&~ 0.13 ~&~ 0.14 ~&~ 0.16 ~&~ 0.10 ~&~ 0.12 \\
  FWW ~&~ 71.58 ~&~ 38.15 ~&~ 47.03 ~&~ 50.28 ~&~ 80.49 ~&~ 132.41 ~&~ 137.05 ~&~ 54.40 ~&~ 43.71 \\
  \hline
\end{tabular}
\vspace{1em}
\caption{Standard Deviation (multiplied by 1000) of DTI and Ball-and-Stick metrics averaged on each vertebral level. Diffusivities are measured in mm$^2/$s.
\label{table:stdev}}
\end{center}
\end{table}

Besides, the reproducibility of DTI and Ball-and-Stick metrics for white matter on controls can be visualized using a Bland-Altman plot \cite{bland1986statistical}. For this analysis, we computed the average of each metric within the white matter of C3-C5, giving a single data point for every metric, subject and scan. The solid red line represents the average difference between scan2 and scan1 and the dashed lines indicate the 95\% confidence interval (CI). All points of controls (except two for AD and FA) for all metrics fall within the 95\% CI, meaning that the studied diffusion metrics are reproducible. They can therefore be further used to detect temporal changes in patients. Results are reported on Fig.~\ref{fig:bland-altman}.

\subsection{Patient-based longitudinal evolution}
\begin{figure}[H] 
  \includegraphics[width=0.95\textwidth]{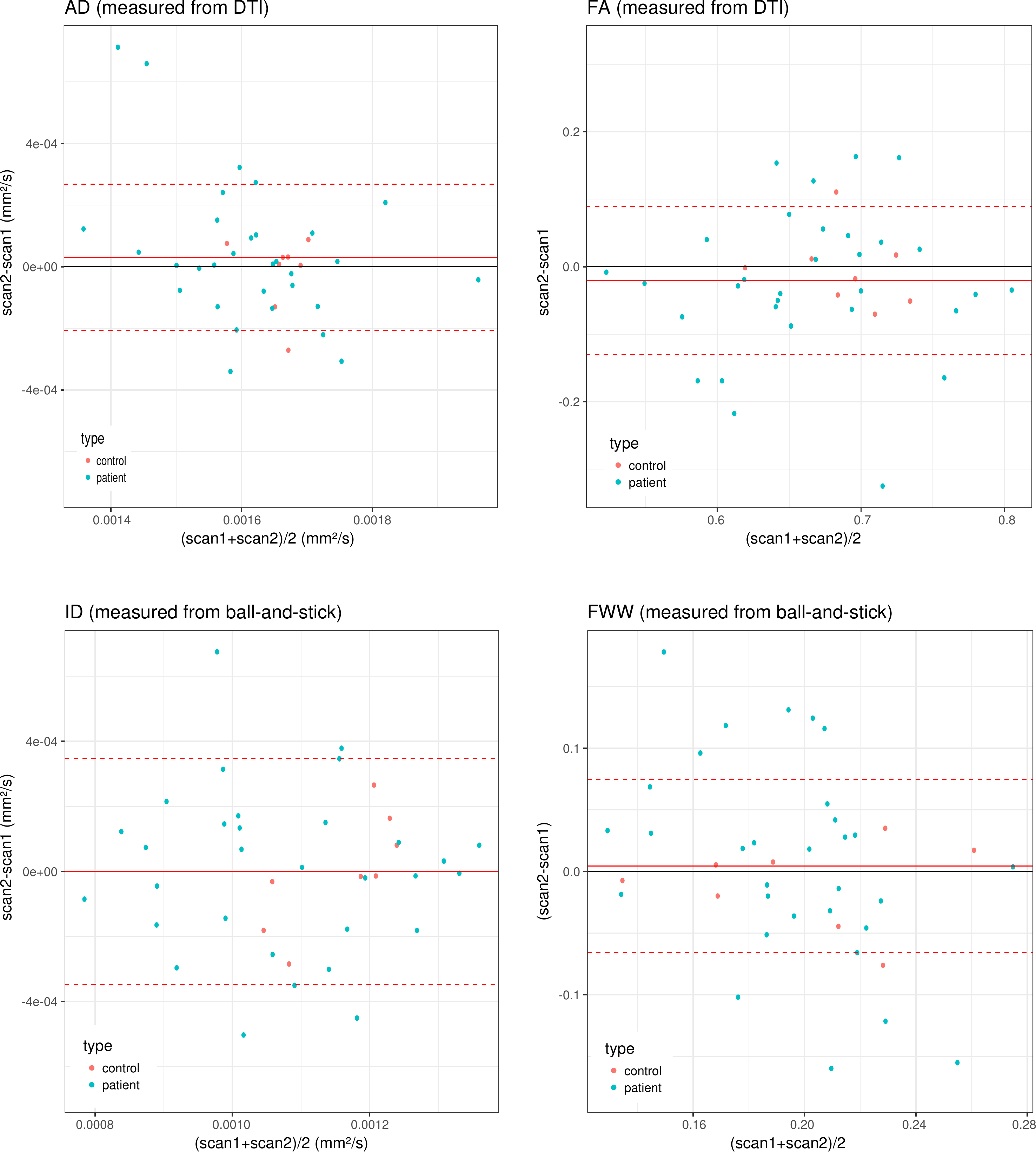}
  \caption{Red: Bland-Altman plot for controls (scan and re-scan) for DTI (top) and ball-and-stick(bottom). Associated confidence interval is represented by the dashed lines. Blue: overlaid metrics difference Scan(M12)-Scan(M0) for patients; data points falling outside the 95\% confidence interval correspond to significant evolution between M0 and M12. \label{fig:bland-altman}}
\end{figure}

The Bland-Altman plot computed on controls defines confidence intervals for each metric averaged on C3-C5. We overlaid on these Bland-Altman plots corresponding values for patients, which allows identification of significant evolution of a given metric between scan and rescan for each patient. In Fig.~\ref{fig:bland-altman}, we can therefore identify significant longitudinal evolution of microstructure-based measures between baseline (M0) and 12 months follow-up (M12). Detailed results are reported on Table~\ref{table:evol_patients} for specific patients, for which several metrics show significant evolution between M0 and M12.

\begin{table}
\begin{center}
\begin{tabular}{c|c|c|c|c|c|c}
  \hline
  \multicolumn{7}{c}{\textbf{M12-M0 for C3C5}}\\
  \hline
  \textbf{Patient} ~&~ Age ~&~ Sex ~&~ ADx$10^{3}$ ~&~ FA ~&~ IDx$10^{3}$ ~&~ FWW \\
  MS Patient-04 ~&~ 32 ~&~ F ~&~ +0.712 ~&~ -0.325 ~&~ NSV ~&~ +0.178\\
  MS Patient-16 ~&~ 31 ~&~ F ~&~ +0.659 ~&~ NSV ~&~ +0.675 ~&~ +0.131\\
  MS Patient-35 ~&~ 22 ~&~ F ~&~ -0.205 ~&~ -0.169 ~&~ NSV ~&~ +0.118\\
  MS Patient-69 ~&~ 36 ~&~ F ~&~ -0.307 ~&~ +0.161 ~&~ NSV ~&~ -0.159\\
  \hline
\end{tabular}
\vspace{1em}
\caption{Evolution of AD, FA (DTI) and ID, FWW (Ball-and-Stick) metrics averaged over C3-C5 vertebral levels between baseline and 12-months follow-up. Diffusivities are measured in mm$^2/$s. NSV: Non Significant Value, referring to Bland-Altman plot.\label{table:evol_patients}}
\end{center}
\end{table}

\section{Discussion}
In Table~\ref{table:evol_patients}, we reported patients for which at least three diffusion metrics evolved significantly between M0 and M12, with respect to the confidence intervals reported in Fig.~\ref{fig:bland-altman}. For patients 04 and 35, we can observe a drop in FA, associated with an increase in the FWW; conversely for patient 69, a increase of FA is associated with a drop in FWW. For these three patients, ID did not change significanlty, which could mean that the change in AD for the DTI model is in fact only due to an increase of the free water compartment, rather than a change in the fibers themselves. Note that for patient 16, no significant change in FA is reported, however there is an increase in the FWW. In general, we observe a complementary between the evolution of metrics extracted from DTI and from Ball-and-Stick.

\section{Conclusion}
In this work, we proposed a framework for studying the evolution of microstructure-related parameters measured with diffusion MRI in the spinal cord white matter of MS patients. Based on a group of healthy controls, we were able to define confidence intervals for diffusion-based metrics for C3-C5 levels in the cervical spine. Using these confidence intervals, we can follow the longitudinal evolution of the same metrics for each patient, and identify abnormal trajectories associated with the pathology. Comparing metrics based on DTI and Ball-and-Stick suggests that both models provide complementary information. This suggests that even for clinical data, multi-compartment models provide novel information about the evolution of tissue microstructure, and should be included in the processing workflow. Future work will include definition of confidence intervals for each vertebral level and study of how the evolution of diffusion MRI indices correlate with clinical scores.

\section*{Acknowledgement}
MRI data acquisition was supported by the Neurinfo MRI research facility from the University of Rennes I. Neurinfo is granted by the European Union (FEDER), the French State, the Brittany Council, Rennes Metropole, Inria, Inserm and the University Hospital of Rennes.

\bibliographystyle{ieeetr}
\bibliography{biblio}

\begin{thebibliography}{10}

\bibitem{barkhof2002clinico}
F.~Barkhof, ``The clinico-radiological paradox in multiple sclerosis
  revisited,'' {\em Current opinion in neurology}, vol.~15, no.~3,
  pp.~239--245, 2002.

\bibitem{cohen2014quantitative}
J.~Cohen-Adad and C.~Wheeler-Kingshott, {\em Quantitative MRI of the spinal
  cord}.
\newblock Academic Press, 2014.

\bibitem{mohammadi2013impact}
S.~Mohammadi, P.~Freund, T.~Feiweier, A.~Curt, and N.~Weiskopf, ``The impact of
  post-processing on spinal cord diffusion tensor imaging,'' {\em Neuroimage},
  vol.~70, pp.~377--385, 2013.

\bibitem{stroman2014current}
P.~W. Stroman, C.~Wheeler-Kingshott, M.~Bacon, J.~Schwab, R.~Bosma, J.~Brooks,
  D.~Cadotte, T.~Carlstedt, O.~Ciccarelli, J.~Cohen-Adad, {\em et~al.}, ``The
  current state-of-the-art of spinal cord imaging: methods,'' {\em Neuroimage},
  vol.~84, pp.~1070--1081, 2014.

\bibitem{clark2000diffusion}
C.~A. Clark, D.~J. Werring, and D.~H. Miller, ``Diffusion imaging of the spinal
  cord in vivo: estimation of the principal diffusivities and application to
  multiple sclerosis,'' {\em Magnetic resonance in medicine}, vol.~43, no.~1,
  pp.~133--138, 2000.

\bibitem{behrens2007probabilistic}
T.~Behrens, H.~J. Berg, S.~Jbabdi, M.~Rushworth, and M.~Woolrich,
  ``Probabilistic diffusion tractography with multiple fibre orientations: What
  can we gain?,'' {\em Neuroimage}, vol.~34, no.~1, pp.~144--155, 2007.

\bibitem{xu2013improved}
J.~Xu, J.~S. Shimony, E.~C. Klawiter, A.~Z. Snyder, K.~Trinkaus, R.~T.
  Naismith, T.~L. Benzinger, A.~H. Cross, and S.-K. Song, ``Improved in vivo
  diffusion tensor imaging of human cervical spinal cord,'' {\em Neuroimage},
  vol.~67, pp.~64--76, 2013.

\bibitem{de2017sct}
B.~De~Leener, S.~L{\'e}vy, S.~M. Dupont, V.~S. Fonov, N.~Stikov, D.~L. Collins,
  V.~Callot, and J.~Cohen-Adad, ``Sct: Spinal cord toolbox, an open-source
  software for processing spinal cord mri data,'' {\em Neuroimage}, vol.~145,
  pp.~24--43, 2017.

\bibitem{ruthotto2012diffeomorphic}
L.~Ruthotto, H.~Kugel, J.~Olesch, B.~Fischer, J.~Modersitzki, M.~Burger, and
  C.~Wolters, ``Diffeomorphic susceptibility artifact correction of
  diffusion-weighted magnetic resonance images,'' {\em Physics in Medicine \&
  Biology}, vol.~57, no.~18, p.~5715, 2012.

\bibitem{snoussi2017comparison}
H.~Snoussi, E.~Caruyer, O.~Commowick, E.~Bannier, and B.~Christian,
  ``Comparison of inhomogeneity distortion correction methods in diffusion mri
  of the spinal cord,'' in {\em ESMRMB-34th Annual Scientific Meeting European
  Society for Magnetic Resonance in Medecine and Biology}, 2017.

\bibitem{de2018pam50}
B.~De~Leener, V.~S. Fonov, D.~L. Collins, V.~Callot, N.~Stikov, and
  J.~Cohen-Adad, ``Pam50: Unbiased multimodal template of the brainstem and
  spinal cord aligned with the icbm152 space,'' {\em NeuroImage}, vol.~165,
  pp.~170--179, 2018.

\bibitem{levy2015white}
S.~L{\'e}vy, M.~Benhamou, C.~Naaman, P.~Rainville, V.~Callot, and
  J.~Cohen-Adad, ``White matter atlas of the human spinal cord with estimation
  of partial volume effect,'' {\em Neuroimage}, vol.~119, pp.~262--271, 2015.

\bibitem{bland1986statistical}
J.~M. Bland and D.~Altman, ``Statistical methods for assessing agreement
  between two methods of clinical measurement,'' {\em The lancet}, vol.~327,
  no.~8476, pp.~307--310, 1986.

\end{thebibliography}

\end{document}